# Unveiling Unconventional Ferroelectric Switching in Multiferroic $Ga_{0.6}Fe_{1.4}O_3$ Thin Films Through Multiscale Electron Microscopy Investigations

*Anna Demchenko[a], Suvidyakumar Homkar[a], Corinne Bouillet[a], Christophe Lefèvre[a], François Roulland[a], Daniele Preziosi[a], Gilles Versini[a], Cédric Leuvrey[a], Philippe Boullay[b,*], Xavier Devaux[c,*], and Nathalie Viart[a,*]*

[a] Université de Strasbourg, CNRS, IPCMS, UMR 7504, 67000 Strasbourg, France
[b] Normandie Université, ENSICAEN, UNICAEN, CNRS, CRISMAT, 14050 Caen, France
[c] Université de Lorraine, CNRS, IJL, 54000 Nancy, France

Corresponding authors: viart@unistra.fr, philippe.boullay@ensicaen.fr, xavier.devaux@univ-lorraine.fr

**Abstract**

Understanding the polarization switching mechanisms at play in ferroelectric materials is crucial for their exploitation in electronic devices. The conventional centrosymmetric reference structure-based mechanism which accounts for ferroelectricity in most of the usual displacive ferroelectric materials is too energy-demanding for some newly diagnosed ferroelectric materials such as the $Ga_{2-x}Fe_xO_3$ ($0.8 \leq x \leq 1.4$) compounds. Some alternative theoretical propositions have been made and need experimental confirmation. A dual-scale electron microscopy study is performed on thin films of the $Ga_{0.6}Fe_{1.4}O_3$ multiferroic compound. A wide scale precession-assisted electron diffraction tomography study first allows the determination of the structure the compound adopts in thin films, and even permits the refinement of the atomic positions within this structure. Cationic mobility is suggested for two of the atomic positions through the existence of extra electronic density. A local *in situ* high resolution scanning transmission electron microscopy study then allows confirming these mobilities by directly spotting the cationic displacements on successively acquired images. The whole study confirms an unconventional switching mechanism *via* local domain wall motion in this compound.

## 1. Introduction

Multiferroic materials showing simultaneous and coupled electric and magnetic orderings are of particularly high technological importance as they open paths towards innovative electronics, in particular in the field of spintronics. Materials presenting both orders at room temperature are however extremely scarce and up to now only bismuth ferrite, $BiFeO_3$ (BFO), has unambiguously met the criteria. It has robust orders, preserved well above room temperature, but is antiferromagnetic, which certainly shadows its applications perspectives, even if, under strain in thin films, its magnetic properties can be enhanced.

An attractive alternate candidate to room temperature multiferroicity is gallium ferrite, $Ga_{2-x}Fe_xO_3$ ($0.8 \leq x \leq 1.4$) (GFO). This material was first addressed in the literature in the 60s [1–4]. In the 2000s, it was however the object of a renewal of interest, building up on the considerable progresses achieved by characterization techniques [5]. GFO crystallizes in the *Pc2₁n* (equivalently *Pna2₁*) space group, different from the perovskite structure commonly adopted by other multiferroic compounds, with $a = 0.8765(2)$ nm, $b = 0.9422(2)$, and $c = 0.5086(2)$ for x=1.4.[6] The structure presents four different cationic sites, a tetrahedral one, labelled Ga1, and three octahedral ones, Ga2, Fe1, and Fe2 (**Figure 1**). The magnetic moments carried by the Fe1 and Ga1 sites, are antiferromagnetically coupled to those carried by the Fe2 and Ga2 sites. A cationic disorder makes the compound ferrimagnetic even for the x=1 composition. The robust ferrimagnetic properties of this compound (Curie temperature above room temperature for x > 1.3 and a room temperature saturation magnetization of 100 emu/cm³, for example, for the x = 1.4 sample) are unquestionably a major advantage when compared to BFO.

The situation concerning the ferroelectric properties of this material is however less clear. An electric polarization of 25 µC/cm$^2$ has been determined for the material from its electronic structure, using first principles methods and the modern theory of polarization [7]. If early works don't address the reversibility of the polarization, recent experimental works have unambiguously shown the ferroelectric character of GFO in thin films [8,9]. The remanent polarization however varies rather strongly from one work to another, from 0.5[8] to 15[9] µC/cm$^2$. The mechanism responsible for the ferroelectric switching is still subject to debate. A first method estimates the electric polarization by following its variation on a path connecting the two opposite polar structures trough a centrosymmetric structure [7,9]. In this approach, the closest centrosymmetric supergroup structure to *Pna2$_1$*, as determined through theoretical group analysis, is a *Pnna* structure. The activation free energy to switch the polarization varies from 0.52 to 1.05 eV per formula unit (f.u.), depending upon the way the calculation is performed [7,9–11]. This energy is about 20 times the energy required to reverse the polarization in one of the mostly used displacive ferroelectrics, Pb(Zr,Ti)O$_3$ [12]. If it can be considered as a positive sign concerning the robustness of polarization in GFO and high ferroelectric transition temperature, it is certainly a major concern when elucidating the actual reversal mechanism. Recent studies identify alternate low-energy polarization switching paths through stochastic surface walking simulation. The studies concern the isostructural compound ε-Fe$_2$O$_3$, for which the unicity of the cationic element type makes the problem easier. They find new possible centrosymmetric reference structures in space groups *Pbcn* or *P2/c*, with much reduced potential barriers of 85 meV[13] and 59 meV[14] per f.u., respectively, when compared to the 340 meV/f.u. calculated for the *Pnna* centrosymmetric reference structure for this compound [13]. The study has been extended to the multi-cationic compound GaFeO$_3$ in reference [13].

The proposed *Pbcn* transition state is not centrosymmetric any more, and the two opposite polarizations are no longer equivalent. GFO can be considered within this frame as an asymmetric multiferroic [15]. The two different energy barriers are then 254 and 105 meV/f.u., and are comparable to what is calculated for BFO [16], which makes this path more realistic than the *Pnna*-based one. However, some even more promising paths are perceived when involving the domain walls in the switching mechanism. The polarization reversal then implies very limited cationic displacements in the vicinity of the domain wall and the energy barrier can be lowered to as low as 22 meV/f.u. in the case of $\varepsilon$-$Fe_2O_3$ [14].

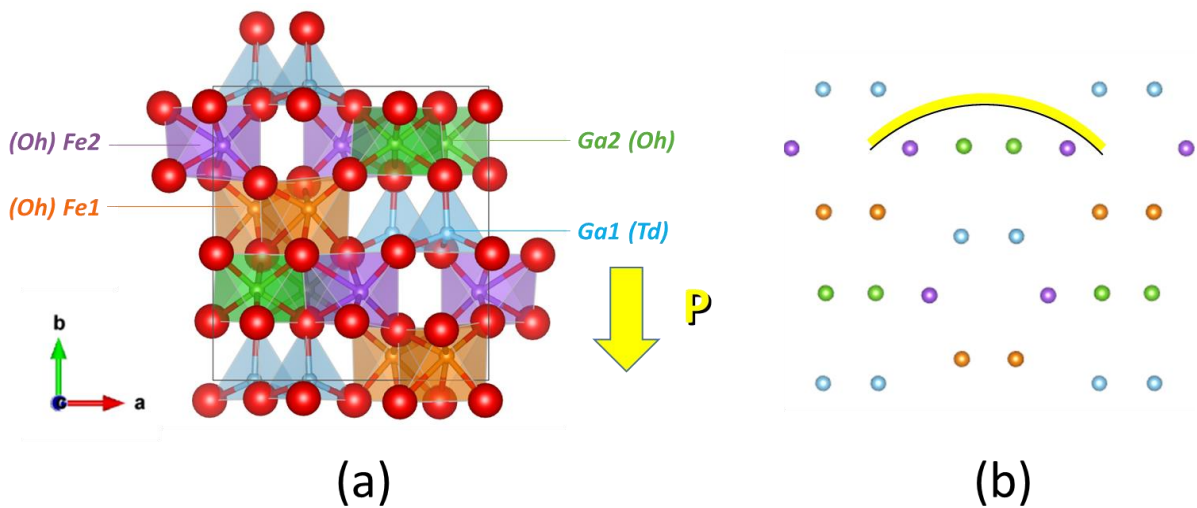

**Fig. 1**. Structural information. a) Unit cell of the orthorhombic $Pc2_1n$ structure of GFO with its four cationic sites. The polarization calculated for this structure is 25 $\mu C/cm^2$, aligned with *b*, but pointing in the opposite direction [7], b) cations-only vision of the GFO structure, highlighting the concave shape adopted by the atoms in the Fe2-Ga2 sites for a downwards polarization. This hallmark allows an easy determination of the polarization orientation in atomic resolved transmission electron micrographs [17].

Various possibilities can therefore be considered for the ferroelectric switching in GFO, from a conventional centrosymmetric reference-based reversal made easier by the crystallization of GFO in a space group different from the one adopts in bulk, to a domain wall-assisted local atomic displacements-based mechanism. In order to get some insight on this mechanism, we undertook a multiscale transmission electron microscopy investigation, in order to, firstly investigate the space group in which GFO crystallizes in thin films by precession-assisted electron diffraction tomography (PEDT), and secondly to consider local mobility-based mechanisms by analysing high resolution scanning transmission electron microscopy (HR STEM) images.

## 2. Experimental details

$Ga_{0.6}Fe_{1.4}O_3$ thin films were grown by pulsed laser ablation of a sintered ceramic target of the same composition. The ceramic target was synthesized by a ceramic method from a mixture of $Fe_2O_3$ and $Ga_2O_3$ oxides (Alrdrich) attrited in ethanol for 6 hours, shaped into a pellet and heated in air to 1400°C for 24 hours. The distance between the target and the substrate in the PLD chamber is 60 mm. The laser wavelength is $\lambda=248$ nm (KrF), its frequency is 10 Hz, and the laser fluence is 2 J/cm$^2$. The deposition is performed in a $O_2:N_2$ atmosphere of 0.1 mbar onto a $SrTiO_3$ (STO) (111) substrate maintained at 900°C.

The composition of the samples is checked by energy dispersive x-ray spectroscopy (EDX) coupled to a scanning electron microscope (JEOL 6700F) operating at 7 kV in order to minimize the influence of the substrate on the spectra.

The structural characterization of the deposited films was first performed by X-ray diffraction (XRD) measurements using a 5-circles Rigaku SmartLab diffractometer equipped with a rotating Cu anode and operating with a K$\alpha_1$ monochromated radiation ($\lambda = 1.54056$ Å).

Transmission electron microscopy (TEM) was then used for the further structural characterizations. Samples were prepared for plane view observations by edge mechanical polishing (Multiprep). Cross section TEM lamellae were prepared by focused ion beam (FIB) at the IEMN facility (D. Troadec, Lille, France). Selected area electron diffraction (SAED) and high resolution transmission electron microscopy (HR TEM) observations were performed on a probe corrected cold field emission gun microscope JEOL 2100F.

PEDT data were collected with a precession angle of 1.2 degrees on a JEOL 2010 (200kV) transmission electron microscope equipped with an upper-mounted Gatan Orius CCD camera. Intensity extractions and data reduction were performed using PETS [18] and the structure solution and refinement was obtained by using JANA2006 [19].

Atomically resolved images were obtained performing high-resolution scanning transmission electron microscopy (HR STEM) using a probe corrected microscope JEOL ARM200F operating at 80 kV. This rather low accelerating voltage value was found to be the optimal one to avoid any beam damage onto GFO during long acquisitions, while still allowing a good resolution for the observations. The displacement of the polarization walls was activated by the scanning electron probe-induced irradiation. The scanned areas were exposed to electron doses of about $2.5 \, 10^5 \, e^-.nm^{-2}.s^{-1}$. Electron energy loss spectroscopy (EELS) spectrum images (SI) were recorded through a Gatan GIF Quantum ER spectrometer with a probe current of about 50 pA and a spatial sampling of 0.03 nm/spectrum. Two EELS-SI were simultaneously recorded: one for the low-loss part, containing the zero loss, and the other one for the core loss part, to allow advanced data post processing (correction of energy drift, multiple scattering corrections). The spectra were recorded with an electron beam having a half-convergence angle of 24 mrad for a half-collection angle of 56 mrad with an energy dispersion of 0.4 eV/channel and a pixel time of 0.005 s. A multivariate statistical analysis software (temDM MSA) was used to improve the quality of the STEM−EELS data by denoising the core-loss SI.

## 3. Results and discussion

### 3.1. Preliminary information: epitaxial relationships and microstructure

An X-ray diffraction pattern of a GFO film grown on STO (111) is given in **Figure 2**a. It only shows the 0k0 reflections of GFO, indicating the *b*-axis orientation of the films (*Pc2$_1$n*). Phi-scans (with Φ between 0 and 360°) were also performed around the STO 313 and GFO 570 reflections to get insight into the in-plane symmetry of the films (**Figure 2**b). They indicate a six-fold symmetry for the GFO films respecting the following epitaxial relationships [600] GFO (010) // [hkl] STO (111) with [hkl] equal to [$\bar{2}$02], [$\bar{2}$20] or [0$\bar{2}$2], in agreement with previous observations [9,20,21].

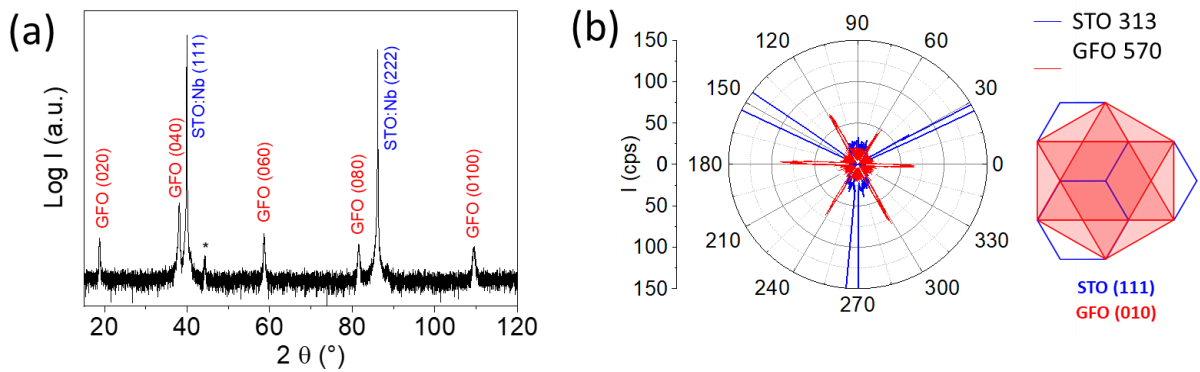

**Fig. 2.** X-ray diffraction characterization of the deposited GFO thin film with (a) θ-2θ scans indicating a (0k0) orientation of GFO (*Pc2$_1$n*) and (b) phi-scans showing three in-plane variants for GFO, in agreement with the epitaxial possibilities offered by the STO (111) substrate.

Further insight into the crystallisation of the films could be obtained from HR-TEM observations of both cross sections and plane views of the film (**Figure 3**). Cross sections (**Figure 3**a and **3**c) confirm the [0k0] growth of the film with an observed interplanar distance of b ≈ 0.943 nm. The growth is columnar with crystallites sizes of ca. 10 nm, as already observed for GFO films grown by PLD [9,22]. The SAED pattern observed for this cross section (**Figure 3**d) corresponds to a [112]$_{STO}$ zone axis (ZA) pattern and shows spots corresponding to one of the three GFO variants, the one in the [001]$_{GFO}$ ZA. The GFO diffraction spots are elongated perpendicularly to the growth, due to the reduced size of the crystallites. The GFO 600 and STO $\bar{2}$20 reflections are not superimposed. This indicates that the GFO film is fully relaxed in-plane and not strained by the STO substrate. The plane views confirm the existence of three in-plane variants and the ca. 10 nm size of the crystallites. The EDS analyses, performed with both SEM and TEM, indicate a Fe/Ga ratio in agreement with a $Ga_{0.6}Fe_{1.4}O_3$ stoichiometry, within the experimental errors.

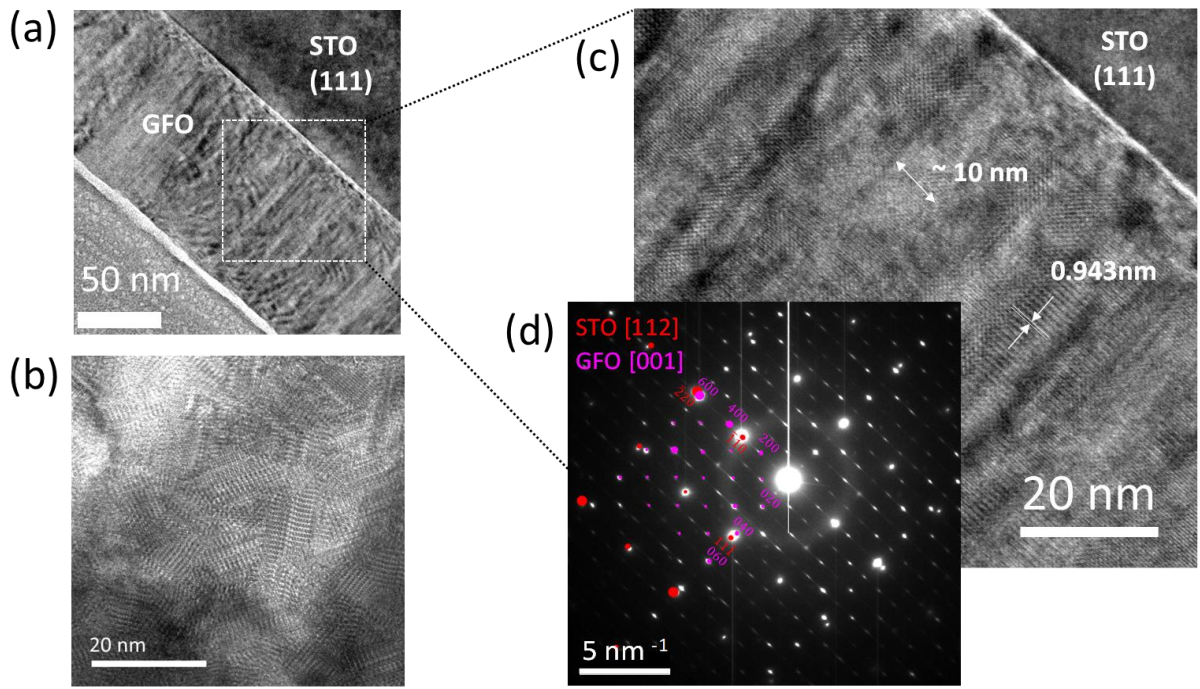

**Fig.3.** HR-TEM observations of the GFO film deposited on a STO(111) substrate a) in cross section, and b) in plane view; c) zoomed image of a) evidencing the size of the crystallites and interplanar distances; d) SAED pattern of the zone observed in c) with spots expected for STO and GFO, in their [112] and [001] zone axes, respectively.

### 3.2. Structural analysis by precession-assisted electron diffraction tomography

The crystallization of GFO in thin films in a space group different from the one adopted in bulk could allow easier paths towards ferroelectricity. The critical point for ferroelectricity in *Pc2$_1$n* GFO is indeed that the energy of the intermediate centrosymmetric supergroup adopted between the two opposite polarisations is too high. If, in thin films, GFO adopts another space group which has a centrosymmetric supergroup of lower energy, this will considerably decrease the switching energy. This would be an explanation for the fact that ferroelectricity has been observed for GFO in thin films. Neutron studies, classically undertaken for the crystallographic study of materials cannot be used for the determination of space groups in thin films because of the too low quantity of matter available. X-ray diffraction is also problematic in the case of thin films, because it has to be performed in reflection mode ; a large part of the reciprocal space is not accessible and the number of measurable reflections is too low to allow unequivocal space group determination [23–25]. This problem no longer exists when considering electron diffraction in a transmission electron microscope and electron diffraction tomography, most often assisted by precession electron diffraction, has proved its efficiency in solving the structure of unknown materials deposited in the form of thin films [26–30]. Nonetheless, in most cases, the structure of the related bulk compound is well known and the challenge is to describe how much the deposited material differs from the bulk reference [31]. Considerable methodological advances have recently been done for the quantitative analysis of the precession-assisted electron diffraction tomography (PEDT) data. Notably, the use of dynamical diffraction theory to refine PEDT data allows overcoming the multiple electron scattering difficulties, and yields access to accurate refined structural models [32–35].

We have undertaken a PEDT study of GFO thin films in order to unambiguously determine the space group in which they crystallize. PEDT data were collected with a precession angle of 1.2 degrees on a JEOL 2010 (200 kV) transmission electron microscope equipped with an upper-mounted Gatan Orius CCD camera. Further details on the experimental procedure used for the acquisition and analysis of the PEDT datasets can be found in a preceding paper of the authors [26]. The area for the data collection was chosen to avoid as much as possible a contribution from the substrate; it was not possible to restrict the collection to only one variant. The reflections collected were used for a 3D reconstruction of the reciprocal lattice bearing in mind the existence of three variants. The cell parameters were obtained for each of them and averaged to a=0.8729(6) nm, b=0.9446(5) nm, c=0.5056(3) nm. A symmetry analysis was then performed through the reconstruction of the acquired slabs of the reciprocal lattice. The indexation was made complex by the presence of the variants and of important diffuse scattering, but decisive information could be obtained on the space groups compatible with the observations (**Figure 4**). Only two space groups happened to be compatible with the observed slabs of the reciprocal lattice: *Pcmn* or *Pc2₁n*. After intensities integration selecting the contribution from one variant, a structural resolution was then conducted using the charge-flipping algorithm SUPERFLIP implemented in JANA2006 [18]. The symmetry analysis performed within this structural resolution allowed to remove the ambiguity concerning the space group and points towards *Pc2₁n* (symmetry agreement factor around 2 for *2₁* instead of 16 for *m*), yielding the precious information that GFO in thin films still adopts its bulk space group. A structural model was then proposed, based on the ten atomic positions of the bulk GFO cell, fours positions for the Ga and Fe cations, with one tetrahedral and three octahedral sites, and six positions for the O ions (see **Table 1**).

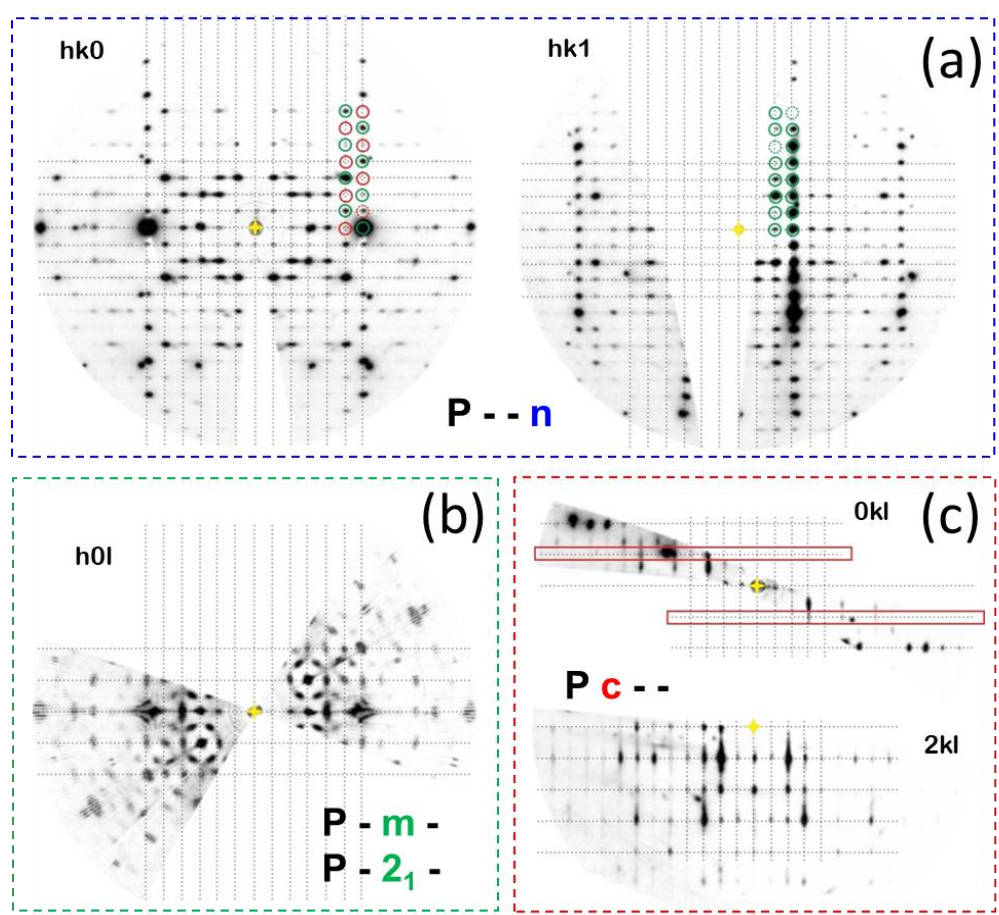

**Fig. 4.** Symmetry analysis performed on various slabs of the reciprocal lattice, allowing to get information on the symmetry elements present in each of the three directions of the cell. The conditions limiting the reflections are found to be: in (a) hk0:h+k=2n, in (b) no condition and in (c) 0kl: l=2n. Two possible space groups result from this analysis: *Pcmn* or *Pc2$_1$n*.

The refinement of this structural model within kinematic approximations yields to a cell which is in good agreement with the expected bulk cell for $Ga_{0.6}Fe_{1.4}O_3$, with Ga in the tetrahedral Ga1 sites and Fe in the three other sites [36](**Figure 5**). From this model, some strong densities (encircled by dashed blue lines in **Figure 5**) are detected in the Fourier difference map at positions intermediate between the Ga1 and Fe1 positions but the reliability factor of the refinement is at this stage however still rather high, with a value of ca. Rw(obs)~40%.

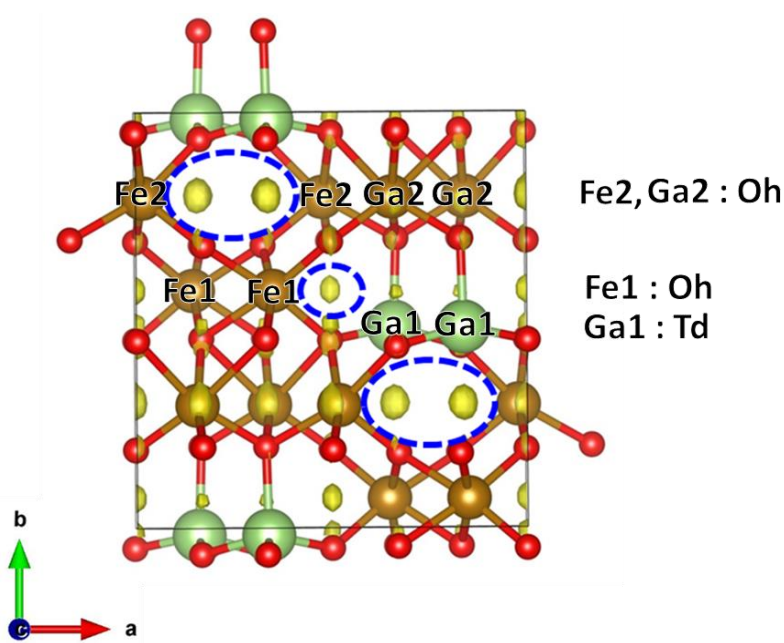

**Fig. 5.** GFO cell obtained from the refinement of a structural model containing the ten atomic positions of bulk GFO, within kinematic approximations (Ga in green, Fe in light brown and O in red). The atomic positions are superimposed with the densities (in yellow) obtained after performing a Fourier difference map calculation. Some extra densities (encircled by a dashed blue line) are visible on positions intermediate between usual Ga1 and Fe1 cationic positions.

Despite the measures taken to render the kinematical approximation as valid as possible for the electron-diffraction data by integrating the diffracted intensities across several beam orientations using precession electron diffraction, the refinements within this approximation yield high figures of merit and questionable accuracy of the refined structure parameters. The situation is even worse here due to the presence of the three orientation variants. In an attempt to get more reliable information from the PEDT data, some refinements were further performed taking into account the dynamical-diffraction effects, according to a method established and tested by Palatinus *et al.* on a wide selection of experimental data [33,34], using the JANA2006 [19] dedicated software. For the present case where three variants contribute simultaneously to the diffracted intensity, we have chosen to perform the refinement on the extracted intensities for only one of these three variants. Doing so we introduce a bias in the analysis because some reflections, common to all three variants, will have an overestimated intensity. We have then separated the reflections common to all the variants (h+l=2n) from those coming only from the selected variant (h+l≠2n) and refined the structure affecting different scales for these 2 sets of reflections. Following this strategy, the dynamical refinement allows to confirm a film structure (**Table 1**) comparable to the bulk GFO structure expected for $Ga_{0.6}Fe_{1.4}O_3$, with Ga in the tetrahedral Ga1 sites and Fe in the three other sites. The observation of Fourier difference maps, after dynamical refinements (**Figure 6**) allows getting maps with much less residual densities compared to **Figure 5**. This suggests that the strong residues, encircled in dashed blue **Figure 5**, were mostly artefacts. In the maps presented **Figure 6**, the residual densities, located in the very close vicinity of the Ga1 and Fe1 sites, would correspond to potential cationic positions in an octahedral or a tetrahedral environment, respectively. Despite the strategy adopted here, we cannot exclude artefacts in difference Fourier maps due to twinning but we believe these residual densities may also be an indicator of some atomic mobility within the cell. The mobility it points to corresponds to the interconversion between tetrahedral Ga1 and octahedral Fe1 sites, which is one way to comprehend the polarization reversal in the GFO structure. To confirm this assumption, we have conducted a structural investigation at a local, atomic, scale using STEM.

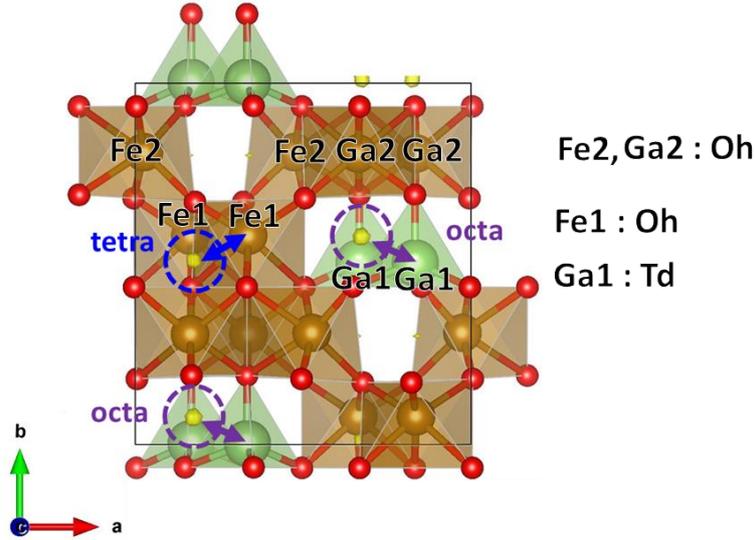

**Fig. 6**. GFO structure obtained from refinements taking into account the dynamical-diffraction effects (Ga in green, Fe in light brown and O in red). The atomic positions are superimposed with the densities (in yellow) obtained after performing a Fourier difference map calculation. Some extra densities (encircled by dashed lines) are visible in the very close vicinity of the tetrahedral Ga1 (purple in octahedral site) and octahedral Fe1 (blue in tetrahedral site) cationic positions.

**Table 1.** Details and results of the GFO thin film structure analysis based on PEDT dynamical refinement. Site occupancy are all equal to 1 and Uiso parameters were all fixed to Uiso(overall) = 0.01 Å².

| | |
|---|---|
| Chemical formula | $Ga_{0.5}Fe_{1.5}O_3$ |
| Temperature (K) | 293 |
| Crystal system, space group | Orthorhombic, $Pc2_1n$ |
| $a$, $b$, $c$ (nm) | 0.8729(6), 0.9446(5), 0.5056(3) |
| Electron wavelength $\lambda$ (Å) | 0.0251 |
| Number of frames | 88 |
| Tilt step (°) | 0.9 |
| Precession angle (°) | 1.2 |
| $\sin(\theta_{max})/\lambda$ (Å$^{-1}$) | 0.7 |
| Completeness (%) | 71.5 |
| No. of measured, observed[$I>3\sigma(I)$] reflections | 4969, 1972 |
| No. of refined parameters, restraints | 191, 0 |
| $g_{max}$ (Å$^{-1}$), $S_{g,max}$ (Å$^{-1}$), $R_{Sg}$, integration steps | 1.7, 0.01, 0.75, 128 |
| $R$(obs), w$R$(obs), w$R$(all), GoF(all) | 0.247, 0.270, 0.277, 11.7 |

| Site label | Atomic occupation | $x$ | $y$ | $z$ |
|---|---|---|---|---|
| Ga1 | Ga | 0.1667(11) | 0 | 0.1812(13) |
| Ga2 | Fe | 0.1620(8) | 0.3022(9) | 0.8265(13) |
| Fe1 | Fe | 0.1637(11) | 0.5750(8) | 0.1711(15) |
| Fe2 | Fe | 0.0244(7) | 0.8056(8) | 0.6655(14) |
| O1 | O | 0.345(3) | 0.441(2) | 0.004(4) |
| O2 | O | 0.497(3) | 0.4357(17) | 0.494(3) |
| O3 | O | 1.009(3) | 0.1793(16) | 0.648(3) |
| O4 | O | 0.170(3) | 0.1876(16) | 0.173(4) |
| O5 | O | 0.173(3) | 0.6693(15) | 0.817(5) |
| O6 | O | 0.160(3) | 0.9372(18) | 0.517(4) |

### 3.3. In situ HR-STEM monitoring of the polarization reversal

STEM annular dark field observations of the GFO thin film in its cross section reveal polarization walls which can be as long as several hundreds of nanometers, crossing tens of crystallites (See **Figure S1** in the Supplementary Information). The orientation of the polarization can be determined on HR-STEM images of GFO crystallites in their (001) ZA, thanks to the hallmark of the Fe2-Ga2 sites: when the four atoms-row in the Fe2-Ga2 sites draw a concave (convex) shape, polarization is downwards (upwards) (see **Figure 1**b). The HR STEM images in **Figure 7**a-d clearly show two distinct zones, in which the polarization is in opposite directions whereas the atomic structure appears to be perfectly continuous and defectless.

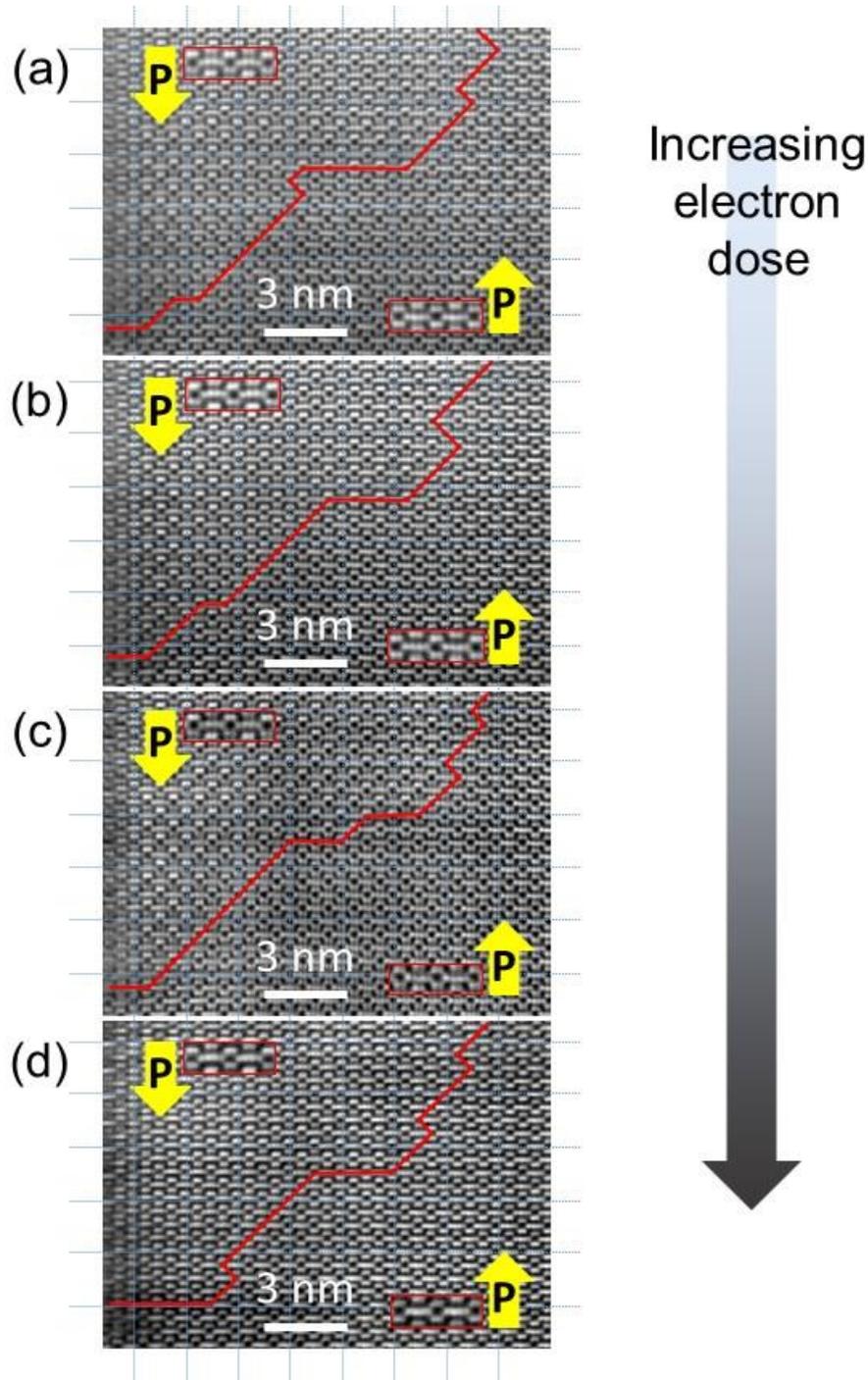

**Fig. 7.** Atomically resolved high angle annular dark-field STEM images of a cross section of a GFO thin film focusing on a zone presenting a polarization wall between two opposite polar domains. The a), b), c) and d) images are acquired successively, and the electron dose increases from a) to d). Between (c) and (d) an EELS SI was recorded on the bottom left part. (see Figure 8). A grid is superposed to the image to facilitate its readability. The horizontal lines of the grid are positioned on the same quadruplets-lines for each image. The red lines indicate in each image the position of the polarization wall reversal. The insets are extracted from the magnification of each image to show the polarization orientation of the atomic network.

The head-to-head polarization domain wall (DW) is highlighted with a red line in all images of **Figure 7**. These images were recorded on the same area over a period of about 2 hours. During this time, the sample was exposed to the electron beam scanning at the same magnification and an image was periodically recorded. To facilitate their comparison, the images were perfectly aligned and superimposed onto a grid. **Figure 7** clearly shows a shift of the DW, with an evolution of the polarization domains with time under the electron dose received by the sample. This DW shift, induced by a local reversal of the polarization, does not seem to be accompanied with any residual mechanical strain. If the contrast in the STEM-high angle annular dark field (HAADF) images tends to indicate that the gallium atoms are positioned in the Td sites for the downward polarization zone (left upper part of the images), it is not strong enough to allow their clear location in the upward polarization area (bottom right). These images do not allow the determination of the cationic distribution in the polarization inversion zone either. To localize gallium and iron on each side of the polarization inversion zone, and to know their respective content in the different sites, a spatially resolved EELS SI was recorded on the lower part of the sample, just after the recording of image 7c. The quantitative chemical maps resulting from the SI processing are presented **Figure 8**. The exact zone onto which the EELS SI was performed is indicated with a yellow frame on the HAADF image (**Figure 8**a). Chemical maps were extracted using the $O_K$, $Fe_L$ and $Ga_L$ signals. The Ga and Fe maps are presented, respectively, in **Figures 8**b and c. On each side of the polarization inversion zone, one can see that the gallium is exclusively localized in the Td sites (Ga1) while the iron is exclusively in the Oh sites (Fe1, Fe2 and Ga2), in accordance with the PEDT analysis. One can notice that the convex or concave shapes of the Fe quadruplets are very easily read on the Fe map, probably even more easily than on the HAADF images. **Figures 8**d,e show enlargements of the chemical maps, focusing on the DW.

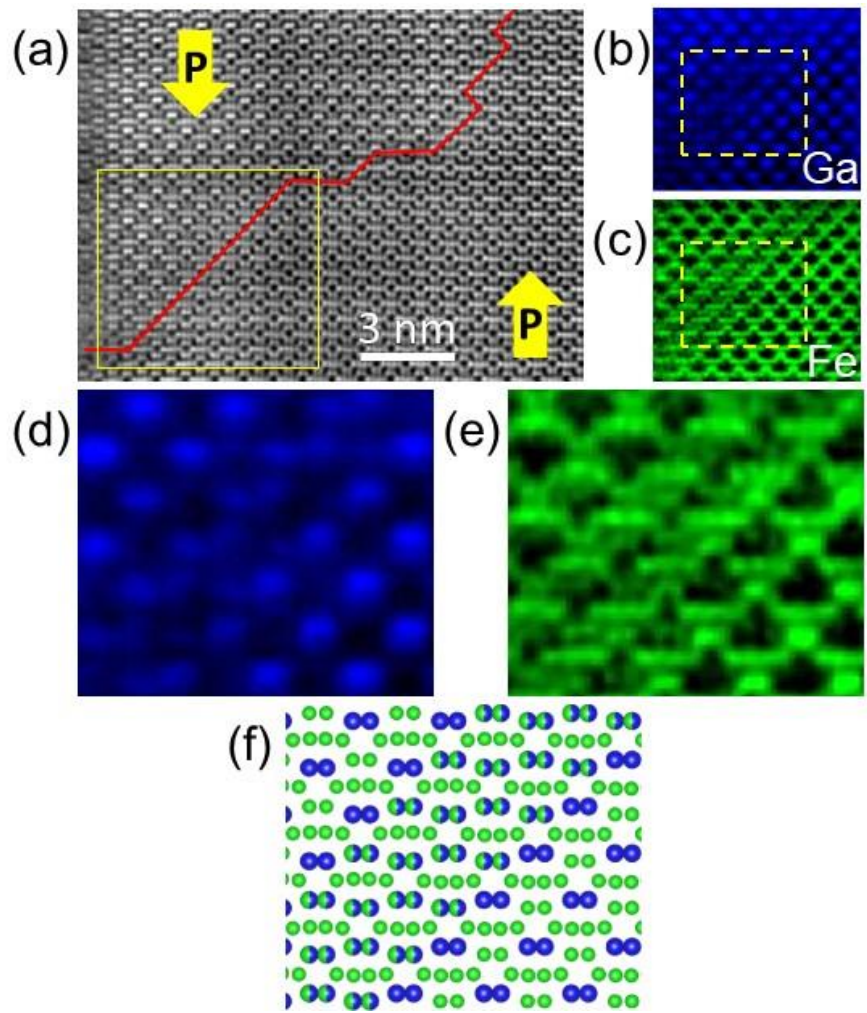

**Fig. 8.** Spatially resolved quantitative chemical maps extracted from EELS SI recorded (a) in the yellow framed area on the HAADF image (same image as in Figure 7c), b) chemical map of gallium, c) chemical map of iron, d) and e) enlargements of images b) and c) to point out the cationic distribution in the DW, f) atomic model showing the Ga and Fe distribution disorder between the Ga1 (Td) and the Fe1 (Oh) sites in the vicinity of the DW.

A Ga and Fe distribution disorder clearly appears in the vicinity of the wall over a thickness of only two unit cells. The atomic model presented in **Figure 8**f depicts the cationic distribution observed in the EELS chemical maps. It is not possible to fully determine the origin of this cationic disorder since the EELS images might result from in-depth atomic column overlapping effects. One could for example imagine a screw-like diffusion mechanism. However, the fact that this cationic disorder is only observed in the vicinity of the domain wall and follows its propagation unambiguously indicates that there is some atomic diffusion.

The additional densities observed at precise locations of the cell in the PEDT study provide supplementary information concerning the cationic mobilities which can be considered, and within that perception, the sites concerned by the cationic mobility are the Fe1 and Ga1 ones.

**Figures 7**a-c show the shift of the polarization wall before the chemical map recording. No trace of this shift is visible on the chemical maps. Another EELS SI was also recorded on the top left after the recording of **Figure 7**d. The same absence of chemical disorder in parts other than the strict wall itself can be observed in the chemical maps associated to this area (see **Figure S2** in the Supplementary Information). There is no trace of cationic disorder outside of the polarization wall, even in the area that was crossed by the wall during its shift. After the wall shift, the reversed area appears with the same cationic distribution as that of the part that has not been affected by the polarization reversal. The shift of the polarization wall that we observed requires the local rearrangement of the oxygen network to transform a Td site into an Oh site. This topotactic transformation necessitates to have some short-range diffusion of Ga and Fe in opposite directions to guarantee the continuity of the structure, the stoichiometry, the absence of stress and the reversal of the polarization. In summary, the cationic disorder observed at the domain wall accompanies its propagation, is only present in its vicinity and nowhere else within the domains. A cationic diffusion therefore has to take place during the propagation of the domain wall to maintain the cationic order observed on each side of the domain wall. This cationic diffusion, involving the local reorganisation of the octahedral (tetrahedral) Fe1 (Ga1) sites into tetrahedral (octahedral) ones perfectly accounts for the polarization reversal. The polarization switching mechanism is therefore underpinned by cationic diffusion evidenced as a cationic disorder in the vicinity of the domain wall.

The two microscopic studies described in this paper, even though they are performed at different scales, therefore converge and shed an important light on the mechanism underlying the polarization reversal in this complex oxide. The observations made here for GFO are in good agreement with those made in its isostructural compound, $\varepsilon$-$Fe_2O_3$ (EFO) [14]. The ferroelectric switching was indeed also observed by HR-STEM to happen through small and near-domain-walls atomic movements. The authors proposed, with the help of a theoretical approach, a domain wall promoted polarization reversal mechanism in which the most implied ions, those which undergo the most important displacements, are in the Fe1 and Ga1 sites. This is exactly what is observed here as well. EFO however only contains one type of cation, the Fe ones. It was therefore impossible to go as far in the understanding of the polarization reversal mechanism as it was the case here thanks to the chemical differentiation allowed by atomically resolved EELS.

## 4. Conclusions

Combining wide scale PEDT studies with local *in situ* HR-STEM observations and spatially resolved EELS, we have been able to elucidate the mechanism at play in the ferroelectric switching of multiferroic $Ga_{0.6}Fe_{1.4}O_4$ thin films. PEDT confirmed that GFO crystallizes in thin films in the same *Pc2₁n* space group as the one observed for bulk, and therefore excluded any conventional mechanism involving a centrosymmetric reference structure-based mechanism, for the theoretical calculations have shown too high energy barriers for these mechanisms. This wide scale study was further exploited thanks to recent progress made in the refinements of the electron diffraction data and allowed obtaining the atomic positions within the unit cell. This evidenced the existence of some mobility in the vicinity of two of the four cationic positions, allowing the polarization reversal, assuming a local small atomic displacements-based mechanism. HR-STEM and STEM-EELS atomically resolved observations allowed confirming this assumption. A cross section zone of the thin film, showing two oppositely oriented polar domains, could be imaged in time, under an increasing electron dose. The downwards polarized domain was observed to gain ground on the upwards one with increasing electron dose and a focus was made on the zones impacted by the polarization reversal. The cationic mobility which was then evidenced is in perfect agreement with the one unveiled by the PEDT study. It establishes that the polarization switching happens in this material through small displacements of atoms in two of the four cationic sites, in the proximity of a domain wall. The whole study therefore provides experimental evidences of the validity of the unconventional local mechanism theoretically proposed for the unicationic $\varepsilon$-$Fe_2O_3$ in the case of the polycationic $Ga_{0.6}Fe_{1.4}O_3$, and even takes the understanding one step further thanks to the chemical differentiation of the cations with atomically resolved EELS.

**Acknowledgements**

This work was funded by the French National Research Agency (ANR) through the ANR MISSION ANR-18-CECE24-0008-01, and within the Interdisciplinary Thematic Institute QMat, as part of the ITI 2021 2028 program of the University of Strasbourg, CNRS and Inserm, it was supported by IdEx Unistra (ANR 10 IDEX 0002) and by SFRI STRAT'US project (ANR 20 SFRI 0012) and ANR-11-LABX-0058_NIE and ANR-17-EURE-0024 under the framework of the French Investments for the Future Program. The authors wish to thank David Troadec (IEMN, Lille, France), Anne-Marie Blanchenet (UMET, Lille, France) and Sylvie Migot (IJL, Nancy, France) for the preparation of the FIB lamellae, as well as the XRD, MEB-CRO, and TEM platforms of the IPCMS. The authors acknowledge financial support from the CNRS-CEA "METSA" French network (FR CNRS 3507) on the platform IRMA (CRISMAT-Caen).

# Supplemental Information

**Unveiling Unconventional Ferroelectric Switching in Multiferroic Ga$_{0.6}$Fe$_{1.4}$O$_4$ Thin Films Through Multiscale Electron Microscopy Investigations**

*Anna Demchenko[a], Suvidyakumar Homkar[a], Philippe Boullay[b,\*], Xavier Devaux[c,\*], Corinne Bouillet[a], Christophe Lefèvre[a], François Roulland[a], Daniele Preziosi[a], Geneviève Pourroy[a], Gilles Versini[a], Cédric Leuvrey[a], and Nathalie Viart[a,\*]*

[a] Université de Strasbourg, CNRS, IPCMS, UMR 7504, F-67000 Strasbourg, France
[b] CRISMAT, Normandie Université, ENSICAEN, UNICAEN, CNRS UMR 6508, 6 Boulevard du Maréchal Juin, Caen, F-14050, France

[c] Université de Lorraine, CNRS, IJL, F-54000 Nancy, France

Corresponding authors: viart@unistra.fr, philippe.boullay@ensicaen.fr, xavier.devaux@univ-lorraine.fr

**SI 1**

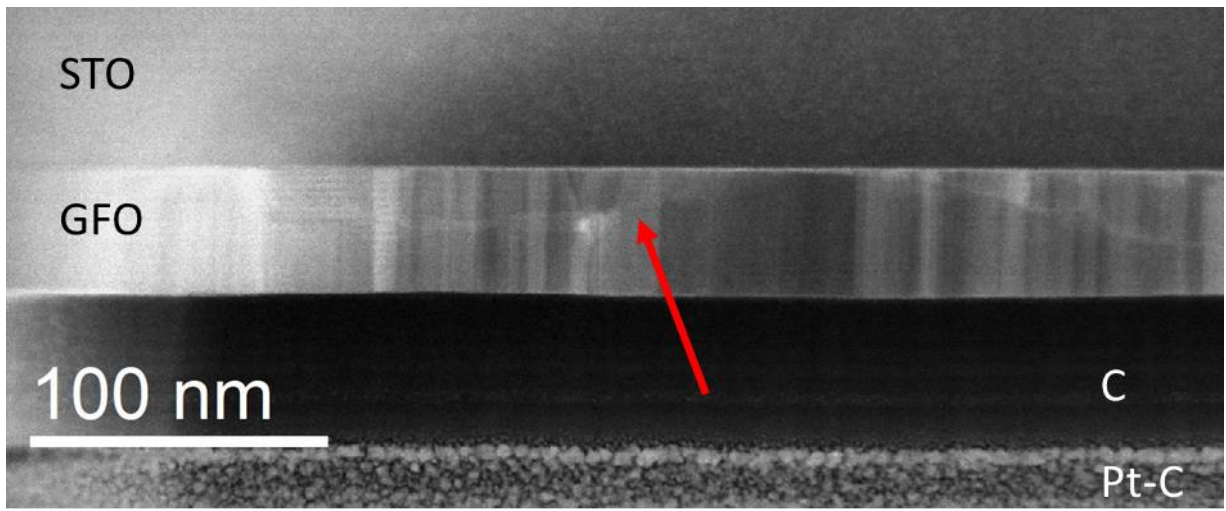

**Figure S1**. Low magnification STEM-ADF image recorded with a collection half-angle range of 30-68 mrad. The arrow points to the area where the images presented in Figures 7 and 8 were recorded.

Polarisation reversal walls, similarly to stacking faults or dislocations, can easily be imaged at low magnification in STEM because a white extra contrast appears on STEM annular dark-field when the area of interest is weakly misoriented as in conventional TEM weak-beam dark-field imaging [1]. For GFO, only HR imaging allows understanding the nature of the defects, and distinguishing an extended stacking fault from a polarisation reversal wall. **Figure S1** shows a low magnification STEM-ADF image of GFO grown onto STO (111). C and Pt layers were deposited on top of the GFO film for FIB processing (visible at the bottom of the image). The white line that crosses the GFO layer from left to right is due to a long polarisation-reversal wall that crosses many columnar grains. The arrow points to the area of the face-to-face polarisation domain wall that was studied at high-resolution by STEM.

**SI 2**

A second spatially resolved ELLS spectrum image (SI) was recorded on the central part of the domain wall (DW), after the recording of the EELS SI presented **Figure 8**, in the area where there was previously a clear shift of the wall (see **Figure 7**). It should be noted that the HAADF survey image presented in **Figure S2**a was also recorded after the one presented in **Figure 8**. No chemical trace of the wall-shift is visible on the chemical maps, as observed for **Figure 8**. The cationic disorder appears only in the DW part oriented at 45° degrees with respect to the growth axis of the layer, as shown in the Ga (Fig. S2b) and Fe (Fig. S2c) maps.

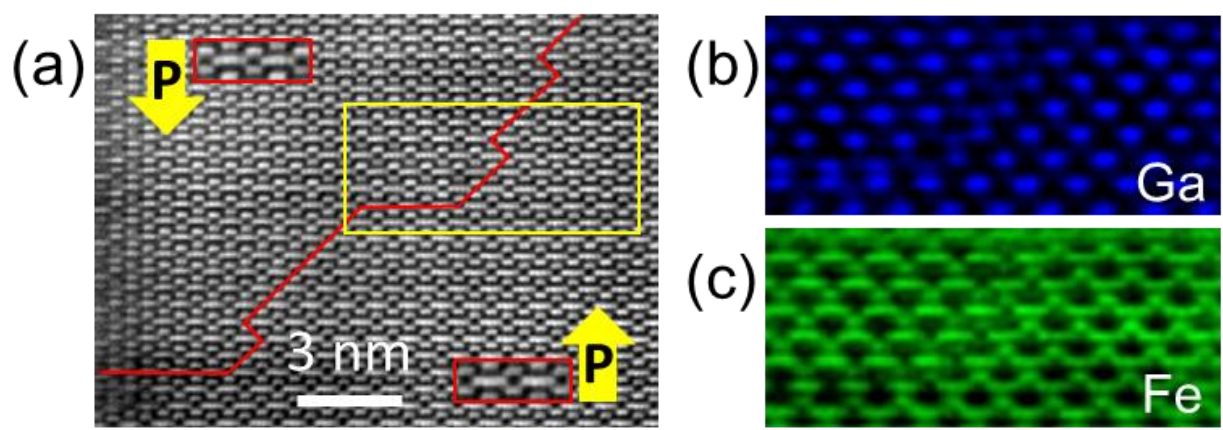

**Figure S2.** a) High resolution STEM-HAADF. The inset magnification of the image shows the local polarisation of the structure. An EELS SI image was recorded in the yellow framed-area. Chemical maps of Ga b) and Fe c) were extracted by processing the EELS-SI.